 \documentclass[final,3p,times,twocolumn]{elsarticle}

 \usepackage{graphicx}

\usepackage{amssymb}
\usepackage{color}

\usepackage{siunitx}
\newcommand{\nuc}[2]{\hbox{$^{#1}$#2}}
\usepackage{gensymb}
\DeclareSIUnit[number-unit-product = {}]{\inchQ}{\textquotedbl}
\DeclareSIUnit[number-unit-product = {\thinspace}]{\inch}{in}

\usepackage{hyperref}

\biboptions{sort,compress}
\journal{Nuclear Instruments and Methods in Physics Research, A}

\begin{document}

\begin{frontmatter}



\title{The CeBrA demonstrator for particle-$\gamma$ coincidence experiments at the FSU Super-Enge Split-Pole Spectrograph}

\author[fsu]{A.L.~Conley}
\address[fsu]{Department of Physics, Florida State University, Tallahassee, FL 32306, USA}

\author[fsu]{B.~Kelly}

\author[fsu]{M.~Spieker\corref{cor1}}
\ead{mspieker@fsu.edu}
\cortext[cor1]{Corresponding author}

\author[fsu]{R.~Aggarwal}

\author[fsu]{S.~Ajayi}

\author[fsu]{L.T.~Baby}

\author[fsu]{S.~Baker}

\author[fsu]{C.~Benetti \fntext[label1]{Current address: Facility for Rare Isotope Beams, Michigan State University, East Lansing, MI 48824, USA} \fnref{label1}}

\author[ursinus]{I.~Conroy}

\author[fsu]{P.D.~Cottle}

\author[davidson]{I.B.~D'Amato}
\address[davidson]{Department of Physics, Davidson College, Davidson, NC 28035, USA}

\author[fsu]{P.~DeRosa}

\author[fsu]{J.~Esparza}

\author[fsu]{S.~Genty}

\author[fsu]{K.~Hanselman \fntext[label2]{Current address: Los Alamos National Laboratory, Los Alamos, New Mexico 87544, USA} \fnref{label2}}

\author[fsu]{I.~Hay}

\author[ursinus]{M.~Heinze}

\author[fsu]{D.~Houlihan}

\author[fsu]{M.I.~Khawaja}

\author[davidson]{P.S.~Kielb}

\author[davidson]{A.N.~Kuchera}

\author[fsu]{G.W.~McCann \fnref{label1}}

\author[fsu]{A.B.~Morelock}

\author[fsu]{E. Lopez-Saavedra}

\author[fsu]{R.~Renom}

\author[ursinus]{L.A.~Riley}
\address[ursinus]{Department of Physics and Astronomy, Ursinus College, Collegeville, PA 19426, USA}

\author[davidson]{G.~Ryan}

\author[fsu]{A.~Sandrik}

\author[fsu]{V.~Sitaraman}

\author[fsu]{E.~Temanson}

\author[fsu]{M.~Wheeler}

\author[fsu]{C.~Wibisono}

\author[fsu]{I.~Wiedenh\"over}

\begin{abstract}
We report on a highly selective experimental setup for particle-$\gamma$ coincidence experiments at the Super-Enge Split-Pole Spectrograph (SE-SPS) of the John D. Fox Superconducting Linear Accelerator Laboratory at Florida State University (FSU) using fast CeBr$_3$ scintillators for $\gamma$-ray detection. Specifically, we report on the results of characterization tests for the first five CeBr$_3$ scintillation detectors of the CeBr$_3$ Array (CeBrA) with respect to energy resolution and timing characteristics. We also present results from the first particle-$\gamma$ coincidence experiments successfully performed with the CeBrA demonstrator and the FSU SE-SPS. We show that with the new setup, $\gamma$-decay branching ratios and particle-$\gamma$ angular correlations can be measured very selectively using narrow excitation energy gates, which are possible thanks to the excellent particle energy resolution of the SE-SPS. In addition, we highlight that nuclear level lifetimes in the nanoseconds regime can be determined by measuring the time difference between particle detection with the SE-SPS focal-plane scintillator and $\gamma$-ray detection with the fast CeBrA detectors. Selective excitation energy gates with the SE-SPS exclude any feeding contributions to these lifetimes.
\end{abstract}

\end{frontmatter}

\section{Introduction}

Particle or $\gamma$-ray spectroscopy experiments alone can oftentimes not provide unambiguous nuclear structure information. For particle spectroscopy, examples could be unresolved excited states due to limited particle energy resolution of the detector or inconclusive angular distributions measured after a nuclear reaction with, {\it e.g.}, a magnetic spectrograph or silicon detector array. For $\gamma$-ray spectroscopy, depending on the design of the experiment and the $\gamma$-ray detectors used, the encountered problems can be quite similar. Particle-$\gamma$ coincidences, when combined with selective nuclear reactions, provide the means to mitigate these problems and also offer a unique window into nuclear-structure phenomena, many of which are important to understanding nuclear reactions taking place in stellar environments. 

Experimental setups which combine high-resolution magnetic spectrographs with $\gamma$-ray detection capabilities are extremely powerful as additional selectivity to different excitation and decay channels can be gained. There are a few examples of existing quasi-permanent setups. They include, {\it e.g.}, the combined GRETINA-S800 \cite{wei17a}, SeGA-S800 \cite{Mue01}, and CAESAR-S800 \cite{wei10a} setups at the National Superconducting Cyclotron Laboratory (NSCL) at Michigan State University \cite{NSCL} [now the Facility for Rare Isotope Beams (FRIB)] as well as the $\gamma$-ray spectrometer DALI2 coupled to the different spectrometers at the RIKEN Radioactive Isotope Beam Factory \cite{dali214a}. These setups are, however, used for experiments with fast rare-isotope beams, where position resolution is critical for Doppler reconstruction. Particle-$\gamma$ coincidence setups with magnetic spectrographs, which temporarily existed at stable beam facilities, include, {\it e.g.}, an HPGe array at the Big-Bite Spectrometer of the KVI Groningen \cite{sav06a} and the CAGRA array at the Grand Raiden spectrometer of the RCNP Osaka \cite{cagra}. First experiments have also been conducted with the BaGeL array at the K600 magnetic spectrometer of iThemba Labs\,\cite{bagel22a}.

In this article, we report on a new experimental setup for particle-$\gamma$ coincidence experiments at the Super-Enge Split-Pole Spectrograph (SE-SPS) of the John D. Fox Superconducting Linear Accelerator Laboratory at Florida State University (FSU) using fast, low-background CeBr$_3$ scintillators for $\gamma$-ray detection. The $\gamma$-ray detection setup at the SE-SPS will be referred to as CeBrA (for CeBr$_3$ Array) in the following. Thanks to the excellent particle energy resolution of the SE-SPS, narrow excitation energy gates can be set, which allow the selective study of the $\gamma$ decay of excited states populated through light-ion induced nuclear reactions with CeBrA. The good energy resolution of the CeBr$_3$ detectors enables the identification of $\gamma$ decays leading to different final states. The narrow excitation energy gates also exclude any feeding contributions when studying particle-$\gamma$ angular correlations for assigning spin-parity quantum numbers and when determining lifetimes with, {\it e.g.}, fast-timing techniques. In contrast to other $\gamma$-ray detectors with comparable or better energy resolution, the CeBr$_3$ detectors can be operated in the intense $\gamma$-ray and neutron background that comes with $(d,p\gamma)$ experiments at rates as high as 250 kilocounts/s (kcps) without running into pile-up problems or losing spectral quality during and after the experiment due to radiation damage.

Detailed studies of the influence of CeBr$_3$ crystal properties on the scintillation process, light yield, energy resolution, timing characteristics, and their implications for spectroscopy applications were first presented in Refs.\,\cite{Qua13a, Fra13a, Qua14a}. Unlike LaBr$_3$, CeBr$_3$ crystals do not suffer from the intrinsic $\gamma$-ray and $\alpha$-particle background between 0.5\,MeV and 3\,MeV, which is caused by the presence of radioactive \nuc{138}{La} and \nuc{227}{Ac} contaminants in LaBr$_3$ crystals \cite{Qua13a}. Because of the lower background, the superiority of CeBr$_3$ over LaBr$_3$ for detecting low-energy $\gamma$ rays emitted in the decay of, {\it e.g.}, actinide nuclei was highlighted in Ref.\,\cite{bnc2}. In addition, the near temperature independence of the scintillation yield for CeBr$_3$ crystals\,\cite{bnc} as well as their very competitive radiation hardness were demonstrated\,\cite{Qua13a, bnc}. Given that the $\sim$\,4\,$\%$ relative energy resolution at 662\,keV is only about 25\,$\%$ worse than for LaBr$_3$ \cite{Qua13a}, CeBr$_3$ detectors offer an attractive and price-effective alternative between low-resolution NaI and comparably expensive, high-resolution HPGe detectors, which also need to be cooled cryogenically and, thus, typically occupy more space around the experimental setup. HPGe detectors are also subject to severe radiation damage in environments with intense neutron background. Furthermore, since the material is composed of heavier elements (the effective $Z$ is 45.9), CeBr$_3$ features a higher $\gamma$-ray detection efficiency than HPGe detectors for comparable crystal sizes. This enhanced detection efficiency is beneficial if higher-energy $\gamma$ rays are to be detected and if a small-scale array is considered. As for LaBr$_3$, it has already been shown that co-doping of the crystal with Strontium can further improve the energy resolution\,\cite{Qua14a}. Large, co-doped CeBr$_3$ crystals are, to our knowledge, not commercially available yet. The timing properties of small, $1'' \times 1''$ CeBr$_3$ detectors have already been studied in a two-detector setup and by replacing one CeBr$_3$ detector with a fast reference BaF$_2$ detector using \nuc{22}{Na} and \nuc{60}{Co} standard calibration sources\,\cite{Fra13a, Qin20a}. The authors of Ref.\,\cite{Fra13a} reported a timing resolution of around 119\,ps when using a fast reference BaF$_2$ detector, which they found to be comparable to LaBr$_3$ detectors making CeBr$_3$ scintillators well suited for fast-timing measurements. Some additional studies of CeBr$_3$ detectors for $\gamma$-ray spectroscopy and lifetime measurements were presented in, {\it e.g.}, Refs.\,\cite{Naq16a, Qin20a, Kab21a, Pep22a, Aog23a, Giovanni_2019}.

This article is structured as follows: Section\,\ref{sec:characterization} will present results for the energy resolution and timing properties of the first five CeBr$_3$ $\gamma$-ray detectors available at FSU. The data were obtained by using standard calibration sources. In Section\,\ref{sec:coincidence}, we will discuss results from the first particle-$\gamma$ coincidence experiments with the CeBrA demonstrator and the SE-SPS. More specifically, it will be shown how particle-$\gamma$ coincidences between the SE-SPS and CeBrA can be used to selectively study the $\gamma$ decays of excited states to different final states and to determine $\gamma$-decay branching ratios. Here, we will feature the data from \nuc{49}{Ti}$(d,p\gamma)$\nuc{50}{Ti} and \nuc{61}{Ni}$(d,p\gamma)$\nuc{62}{Ni} test experiments. The data from the second set of test experiments, \nuc{52}{Cr}$(d,p\gamma)$\nuc{53}{Cr} and \nuc{34}{S}$(d,p\gamma)$\nuc{35}{S}, will be used to highlight the combined setup's ability to measure particle-$\gamma$ angular correlations for spin-parity assignments and to determine nuclear level lifetimes via fast-timing techniques without any feeding contributions.      


\section{Characterization of CeBr$_3$ detectors with standard calibration sources}

\label{sec:characterization}

\begin{figure}
    \centering
    \includegraphics[width=1\linewidth]{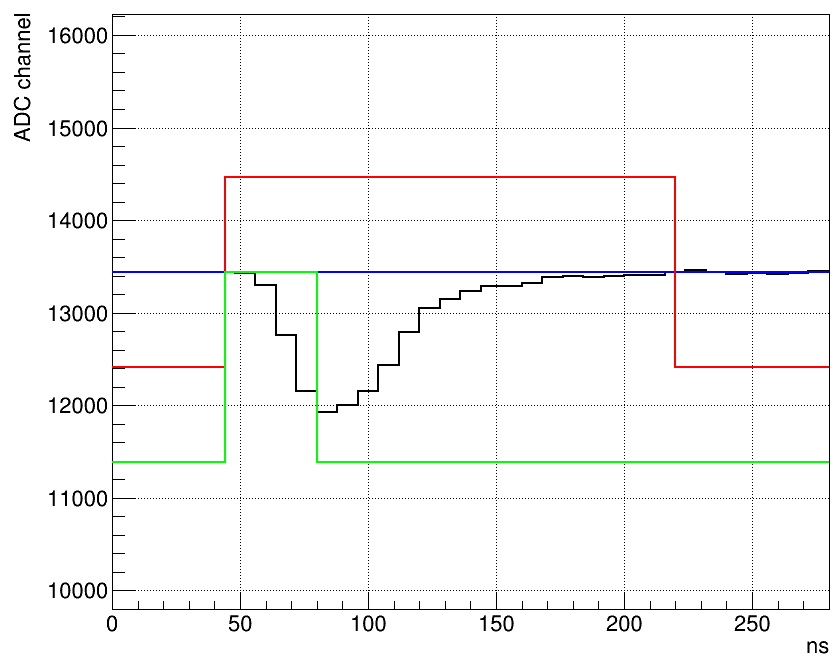}
    \caption{The digital QDC filter windows used for the $2'' \times 2''$ CeBr$_3$ detectors with the CAEN V1725S digitizer using the proprietary DPP-PSD firmware.  The blue line represents the baseline of the signal, while the green and red lines correspond to the short and long gates, respectively. The black line is the detector signal.}
    \label{fig:signal}
\end{figure}

\begin{figure}[t]
    \centering
    \includegraphics[width=1\linewidth]{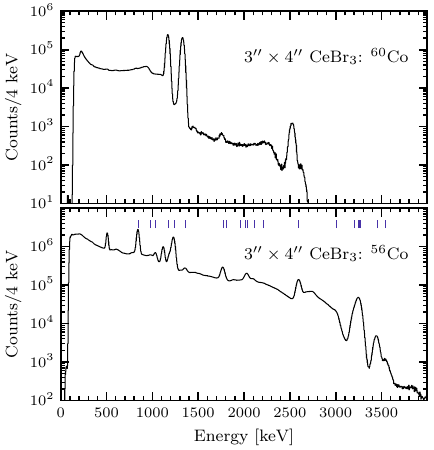}
    \caption{\label{fig:60Co_56Co_spectrum}{Spectrum of $\gamma$ rays emitted after the decay of \nuc{60}{Co} to \nuc{60}{Ni} (top) and \nuc{56}{Co} to \nuc{56}{Fe} (bottom) measured with a $3'' \times 4''$ CeBr$_3$ detector. The bottom figure includes vertical lines to illustrate the $\gamma$ rays associated with the decay of \nuc{56}{Co}.}}
\end{figure}

For particle-$\gamma$ coincidence experiments at the FSU SE-SPS, four $2'' \times 2''$ and one $3'' \times 4''$ low-background, cylindrical CeBr$_3$ detectors with Hamamatsu R6231 and R6233 PMTs (all with positive bias), respectively, were acquired from Berkeley Nucleonics (BNC) \cite{bnc} and Advatech UK \cite{advatech}. Afterwards, they were thoroughly tested. All detectors have a magnetic shield and built-in voltage divider. High voltage is supplied to the detectors using the iseg MPOD HV system with the EHS F020p unit. Detector signals are recorded using a digital data acquisition (DAQ) based on the CAEN V1725S digitizer with the proprietary DPP-PSD firmware \cite{caen}. The digitizer is readout via an optical fiber link connection and data recorded to disk.
 
\subsection{Energy resolution}

\begin{figure}[t]
\centering
\includegraphics[width=1\linewidth]{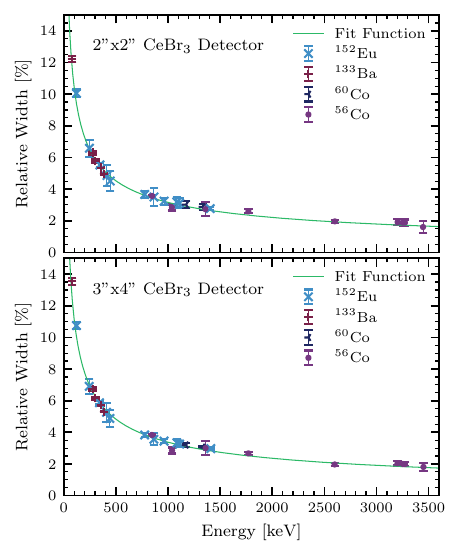}
        \caption{\label{fig:cebr3_width}{The relative full width at half maximum (FWHM) as a function of $\gamma$-ray energy for a $2'' \times 2''$ CeBr$_3$ detector with Hamamatsu R6231 PMT and a $3'' \times 4''$ CeBr$_3$ detector with Hamamatsu R6233 PMT, both with positive bias as recorded with a CAEN V1725S digitizer at FSU. Different standard calibration sources were used.}}
 \end{figure}

The DPP-PSD filter settings described in Ref.\,\cite{caen_dpp} were varied to optimize energy resolution. 
For the $2'' \times 2''$ CeBr$_3$ detectors, the digital QDC filter settings were a gate of 176\,ns, a short gate of 28\,ns, and a pre-gate of 148\,ns. Similarly, for the $3'' \times 4''$ CeBr$_3$ detector, the filter settings were a gate of 200\,ns, a short gate of 28\,ns, and a pre-gate of 148\,ns. An example of the detector signal and the digital QDC filter windows used for the $2'' \times 2''$ CeBr$_3$ detectors is shown in Fig.\,\ref{fig:signal}. Recorded $\gamma$-ray spectra for the $3'' \times 4''$ CeBr$_3$ detector are presented in Fig.\,\ref{fig:60Co_56Co_spectrum}. For both crysal sizes, the energy resolution was found to be equal to the vendors' specification. As expected (see also Ref. \cite{Qua13a}), the full width at half maximum (FWHM) evolves with $1/\sqrt{E_{\gamma}}$ (see Fig.\,\ref{fig:cebr3_width}). The energy resolution of the $3'' \times 4''$ crystal is slightly worse than the one of the $2'' \times 2''$ crystal ($\mathrm{FWHM} = 2.8$\,$\%$ (38 keV) compared to 2.6\,$\%$ (35 keV) at $E_{\gamma}=1.3$\,MeV). Our results are in agreement with the previously reported energy resolution of $2'' \times 2''$ CeBr$_3$ detectors\,\cite{Qua13a}.

\subsection{Timing properties}

\begin{figure}[t]
\centering
\includegraphics[width=.9\linewidth]{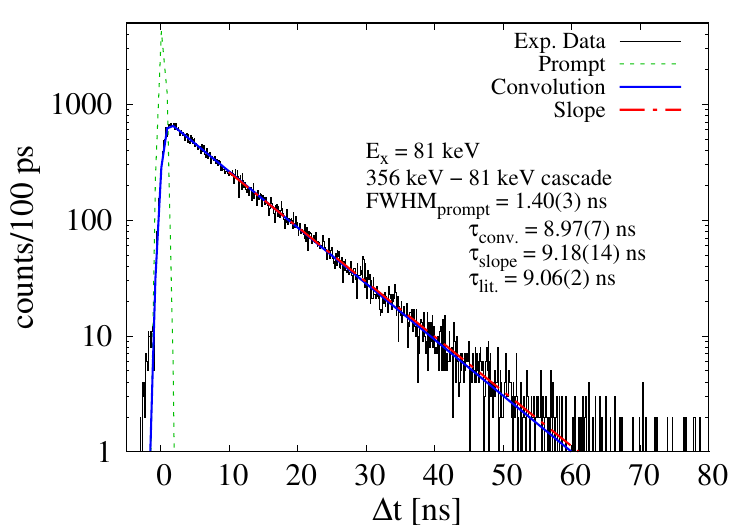}
\includegraphics[width=.9\linewidth]{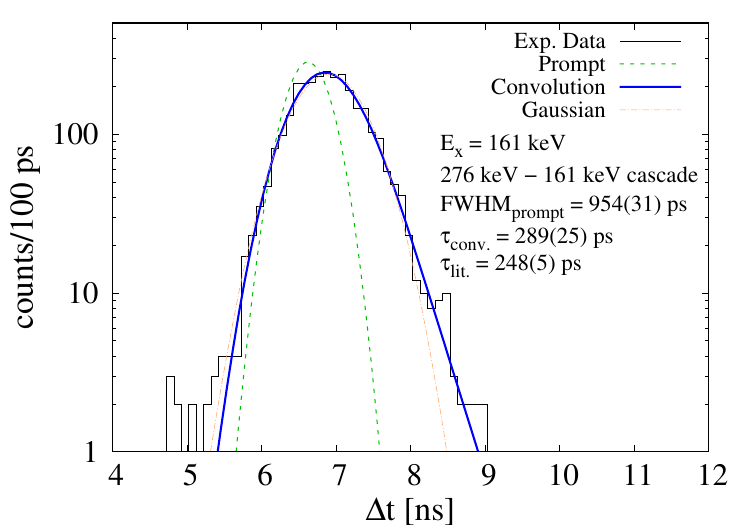}
\caption{Lifetime determination of the 81-keV (top) and 161-keV (bottom) states in \nuc{133}{Cs} with the $2'' \times 2''$ CeBr$_3$ detectors via the 356-keV and 276-keV $\gamma$-decay branches, respectively, which feed the states directly. See text for further details}
\label{fig:lifetime}
\end{figure} 

As mentioned in the introduction, the timing properties of $1'' \times 1''$ CeBr$_3$ detectors have already been studied in a two-detector setup and by replacing one CeBr$_3$ detector with a fast reference BaF$_2$ detector using \nuc{22}{Na} and \nuc{60}{Co} standard calibration sources\,\cite{Fra13a, Qin20a}. The authors of Ref.\,\cite{Fra13a} reported a timing resolution of around 119\,ps measured against a fast reference BaF$_2$ detector, which they found to be comparable to LaBr$_3$(Ce) detectors. No information on the timing resolution was previously available for the $2'' \times 2''$ and $3'' \times 4''$ crystal sizes. For our studies, we used the built-in digital constant fraction discriminator (CFD) offered by CAEN for the V1725S digitizer with the DPP-PSD firmware \cite{caen_dpp} and followed the optimization procedure outlined in, {\it e.g.}, Refs.\,\cite{Qin20a, caen_cfd}. Furthermore, we used a homogeneous setup, i.e., detectors of the same crystal type. To exclude possible DAQ limitations, two $1.5'' \times 1.5''$ LaBr$_3$(Ce) detectors were lent from Michigan State University. In agreement with published results \cite{long19a}, the timing resolution for the two prompt $\gamma$-ray transitions observed after the $\beta^-$ decay of \nuc{60}{Co} was determined to be $\sim 300$\,ps for a setup consisting of two LaBr$_3$(Ce) detectors. For a setup consisting of two $2'' \times 2''$ CeBr$_3$ detectors, we measured a timing resolution between 500\,ps and 590\,ps using the same two prompt $\gamma$-ray transitions. The optimal CFD settings were found to be 44 ns for the CFD delay and 75\% for the CFD fraction, which given the slightly longer rise time of our detectors is in line with the results of Ref.\,\cite{caen_cfd}. The Advatech UK detectors showed slightly better timing properties than the ones we acquired from Berkeley Nucleonics even though the same components are used. For the $3'' \times 4''$ detector, we can only state a prompt timing resolution measured relative to a $2'' \times 2''$ CeBr$_3$ detector. Here, we found a timing resolution of $\sim 750$\,ps between the two prompt lines emitted in the $\beta^-$ decay of \nuc{60}{Co}. The CFD settings used for the $3'' \times 4''$ detector were 68 ns for the CFD delay and 75\% for the CFD fraction. The worse timing resolution of the CeBr$_3$ detectors compared to previous measurements with LaBr$_3$ and smaller-size CeBr$_3$ detectors \cite{Qin20a} can be partly explained due to material differences between LaBr$_3$ and CeBr$_3$ \cite{Fra13a} as well as the larger crystal geometry of our detectors. The Hamamatsu R6231 PMT is also more optimized for energy than for timing resolution (see, {\it e.g.}, Ref. \cite{ved15a}).

After benchmarking the timing resolution, we tested that we could reliably extract lifetimes with the $2'' \times 2''$ CeBr$_3$ detectors. Two examples using the slope and convolution method for previously known lifetimes of excited states in \nuc{133}{Cs} populated after the decay of \nuc{133}{Ba} \cite{ENSDF} are shown in Fig.\,\ref{fig:lifetime}. The lifetime for the 81-keV state determined via the slope and convolution methods is in excellent agreement with the literature value\,\cite{ENSDF}. A slight discrepancy is observed for the shorter lifetime with respect to the adopted value (see Fig.\,\ref{fig:lifetime}). It should be noted, however, that our result for the lifetime of the 161-keV state nicely agrees with other published data, see, {\it e.g.}, \cite{ENSDF,Dey91a}. For comparison, a Gaussian distribution (orange dashed line) has been fitted to the timing distribution observed for the 276 keV -- 161 keV $\gamma$-ray cascade. The asymmetry in the tail indicative of a lifetime in the hundreds of picosecond range can be clearly seen. Note that the FWHM of the prompt contribution changes with $\gamma$-ray energy (see also, {\it e.g.}, Ref.\,\cite{Qin20a, regis11a}).

\section{Commissioning particle-$\gamma$ coincidence experiments with the CeBrA demonstrator at the FSU SE-SPS}

\label{sec:coincidence}

\begin{figure}[t]
    \centering
    \includegraphics[width=1\linewidth]{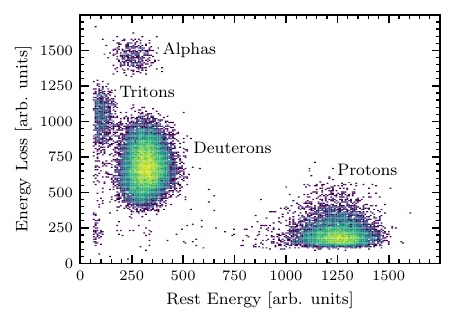}
    \caption{\label{fig:ede}{Particle identification with the SE-SPS focal-plane detector. 
    A 16-MeV deuteron beam impinged onto a self-supporting 413-$\mu$g/cm$^2$ \nuc{49}{Ti} target. Four particle groups corresponding to protons, deuterons, tritons, and alpha particles can be clearly distinguished. The magnetic field of the SE-SPS was set to 8.8 kG and the gas pressure in the focal-plane detector was 160\,Torr.}}
    \label{fig:PID}
\end{figure}

\subsection{The Super-Enge Split-Pole Spectrograph (SE-SPS)}

The Super-Enge Split-Pole Spectrograph (SE-SPS) was moved to FSU after the Wright Nuclear Structure Laboratory (WNSL) at Yale University ceased operation. Like any spectrograph of the split-pole design \cite{Eng79a, SPENCER1967181}, the SE-SPS consists of two pole sections used to momentum-analyze light-ion reaction products and focus them at the magnetic focal plane to identify nuclear reactions and excited states. The split-pole design allows to accomplish approximate transverse focusing as well as to maintain second-order corrections in the polar angle $\theta$ and azimuthal angle $\phi$, i.e., $(x/\theta^2) \approx 0$ and $(x/\phi^2) \approx 0$, over the entire horizontal range \cite{Eng79a, SPENCER1967181}. H. Enge specifically designed the SE-SPS spectrograph as a large-acceptance modification to the traditional split-pole design for the WNSL. The increase in solid angle from 2.8 to 12.8 msr was achieved by doubling the pole-gap, making the SE-SPS well-suited for coincidence experiments. At FSU, the SE-SPS was commissioned in 2018. The Silicon Array for Branching Ratio Experiments (SABRE) was commissioned as the first ancillary detector at the SE-SPS for studying unbound resonances relevant for Nuclear Astrophysics\,\cite{sabre21a}.

 \begin{figure}[t]
\centering
\includegraphics[width=1\linewidth]{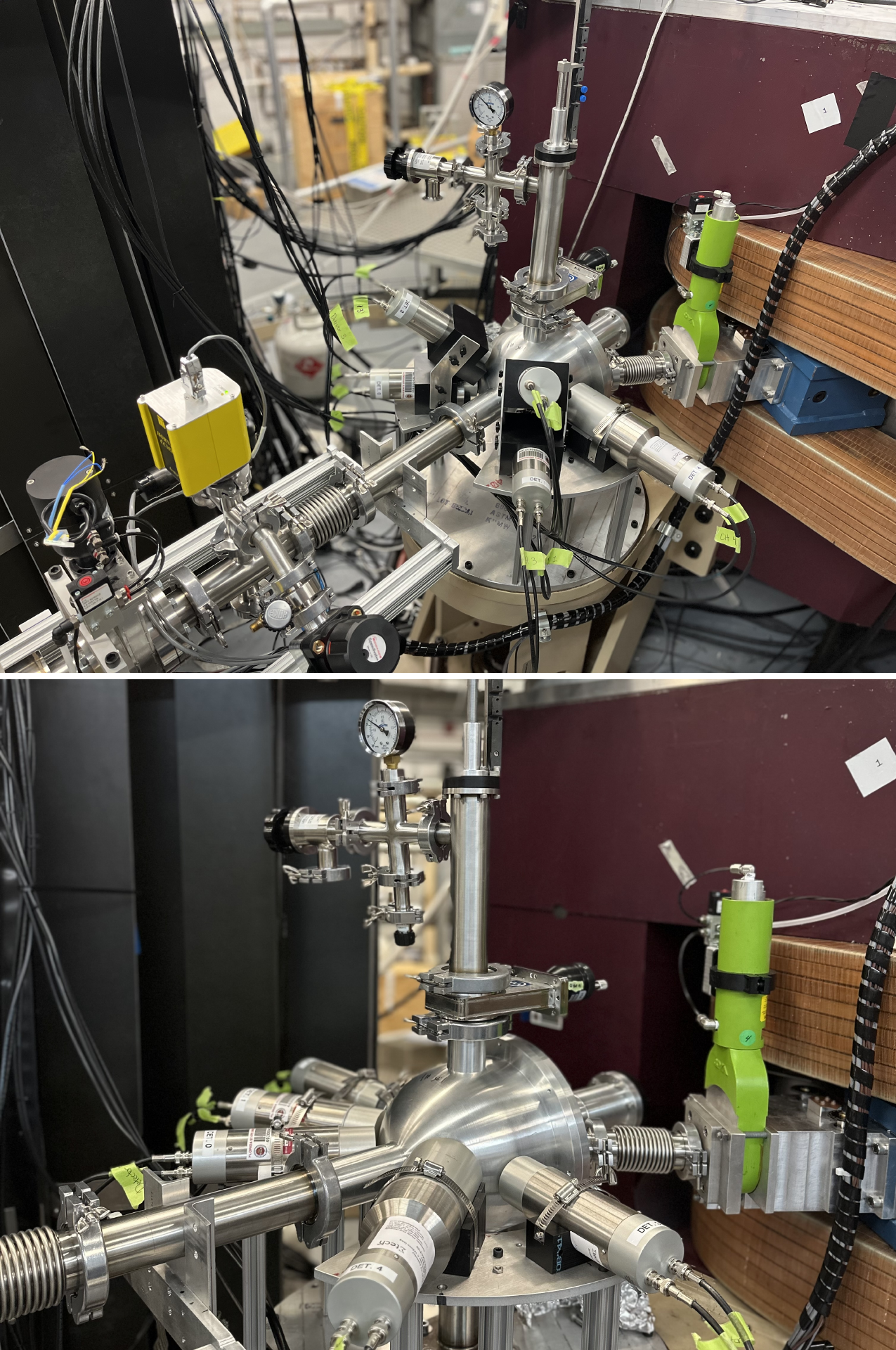}
\caption{\label{fig:cebra}{The first five CeBrA detectors at the FSU SE-SPS around the newly built aluminum scattering chamber. Currently, four $2'' \times 2''$ and one $3'' \times 4''$ CeBr$_3$ scintillators are available for $\gamma$-ray detection. (Top) Configuration for the \nuc{49}{Ti}$(d,p\gamma)$\nuc{50}{Ti} and \nuc{61}{Ni}$(d,p\gamma)$\nuc{62}{Ni} experiments. (Bottom) Configuration for the \nuc{52}{Cr}$(d,p\gamma)$\nuc{53}{Cr} and \nuc{34}{S}$(d,p\gamma)$\nuc{35}{S} experiments. The SE-SPS and its pole faces can be seen to the right of the CeBrA scattering chamber. The entrance to the SE-SPS with its slit box is behind the green gate valve which can be seen in the photos.}} 
 \end{figure}

In singles experiments, i.e., stand-alone mode, the SE-SPS with its present light-ion focal plane detection system (see, {\it e.g.}, Ref.\,\cite{goo20a} for some details) can be used to study the population of excited states in light-ion induced reactions, determine (differential) cross sections and measure the corresponding angular distributions. The focal-plane detector consists of a position-sensitive ionization chamber with two anode wires to measure energy loss in the isobutane gas and a large plastic scintillator to determine the rest energy of the residual particles passing through the detector. A sample particle identification plot with the energy loss measured by the rear anode wire and the rest energy measured by the scintillator is shown in Fig.\,\ref{fig:ede}. Unambiguous particle identification is achieved. Under favorable conditions, the detector can be operated at rates as high as two kilocounts/s (kcps). Examples of angular distributions measured with the SE-SPS in \nuc{50}{Ti}$(d,p)$\nuc{51}{Ti}, \nuc{54}{Fe}$(d,p)$\nuc{55}{Fe}, and \nuc{61}{Ni}$(d,p)$\nuc{62}{Ni} are shown in Refs.\,\cite{Ril21a, Ril22a, Spi23a}. They provide direct information on the angular momentum, $l$, transfer and the involved single-particle levels. Sample position spectra measured with the delay lines are also shown in Refs.\,\cite{Ril21a, Ril22a, Spi23a}. As the energy resolution depends on the solid-angle acceptance, target thickness and beam-spot size, it may vary from experiment to experiment. In standard operation and with a global kinematic correction, a FWHM of 30-50\,keV is routinely achieved. This resolution can be improved further with position-dependent offline corrections.

\begin{figure}[t]
    \centering
    \includegraphics[width=\linewidth]{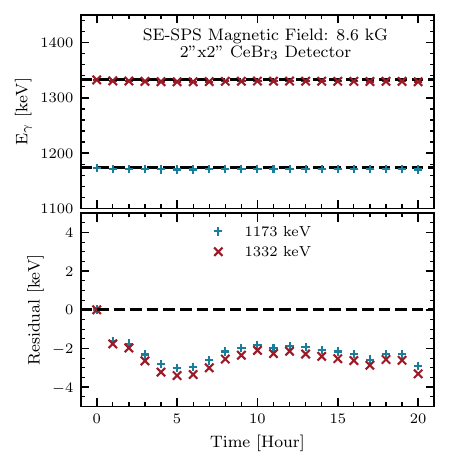}
    \caption{\label{fig:gain_stability}Gain stability study for a $2'' \times 2$'' CeBr$_3$ detector using the 1173-keV and 1332-keV $\gamma$-ray transitions of a \nuc{60}{Co} standard calibration source. The detector was positioned around the CeBrA scattering chamber in front of the SE-SPS with the magnetic field set to 8.6 kG (see Fig.\,\ref{fig:cebra}). See text for discussion.}
\end{figure}

\subsection{New scattering chamber and general setup design}

A new scattering chamber for particle-$\gamma$ coincidence experiments was designed and constructed at FSU. The 8-inch-diameter chamber is made out of aluminum, features 0.2-inch thin walls, the possibility to shield the Faraday cup (beam dump) and to evacuate the target ladder separately. In order to further reduce beam-induced $\gamma$-ray background coming from the Faraday cup, the latter can be shielded with lead. The front piece can be adjusted to the needs of specific experiments, i.e., the scattering angle can be changed by using a different front piece. The initial version has a port to the SE-SPS at $\sim 37$ degrees (measured clockwise relative to the beam axis), a zero-degree port with a Faraday cup, and an additional port for an optional silicon beam-monitor detector. Each chamber-detector distance can be adjusted. For all detectors, the closest distance to the target is 4.5 inches. The adjustable distance of each detector to the chamber allows, in principal, for the addition of shields and attenuators if needed. Fig.\,\ref{fig:cebra} shows the CeBrA demonstrator and scattering chamber in front of the FSU SE-SPS for the two different configurations, which were used for the $(d,p\gamma)$ experiments featured below.

\subsection{Gain stability of CeBr$_3$ detectors at the SE-SPS}

\begin{figure}[t]
    \centering
    \includegraphics[width=\linewidth]{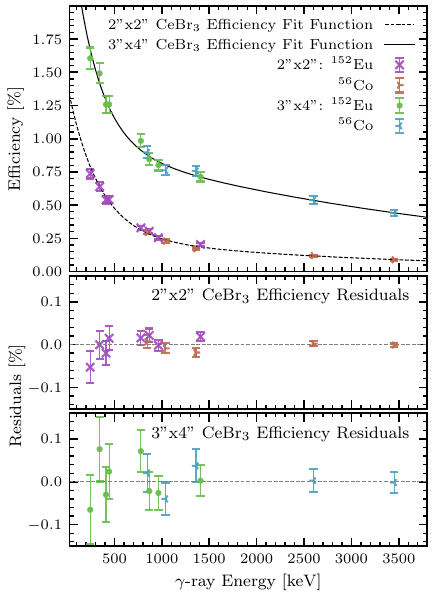}
    \caption{\label{fig:efficiency}(Top) Absolute full-energy peak (FEP) efficiency for a $2'' \times 2$'' and $3'' \times 4''$ CeBr$_3$ detector measured for the second CeBrA configuration shown in Fig.\,\ref{fig:cebra} with a \nuc{152}{Eu} standard calibration source of known activity and a \nuc{56}{Co} source of unknown activity. In overlap regions, the \nuc{56}{Co} data were scaled to the \nuc{152}{Eu} data to obtain absolute FEP efficiencies up to $E_{\gamma} = 3.5$\,MeV. A double exponential function was fitted to the data in both cases. The residuals between the fit and data are shown in the middle and bottom panel, respectively.}
\end{figure}

With the CeBr$_3$ detectors positioned around the CeBrA scattering chamber, we explored the gain stability of the detectors, aiming to understand the potential impact of SE-SPS fringe fields on the PMTs during experiments. The spectrograph was set to a magnetic field strength of 8.6 kG and and the CeBr$_3$ detectors were calibrated using a \nuc{60}{Co} standard calibration source after the first hour-long run (corresponding to hour 0 in Fig.\,\ref{fig:gain_stability}). This energy calibration was then consistently applied to the data collected in subsequent runs and not changed in order to identify possible gain shifts. As illustrated in Fig.\,\ref{fig:gain_stability}, a small initial shift of approximately 2 keV is observed after the first run. Nevertheless, following this initial shift, the PMT gain exhibited predominantly stable behavior throughout the remaining data collection. Given the energy resolution of $\sim 3$\,$\%$ for the $2'' \times 2''$ CeBr$_3$ detectors at these $\gamma$-ray energies, we conclude that this level of gain shift is not significant and can be easily corrected for offline, which we successfully did for the $(d,p\gamma)$ experiments discussed below.

\subsection{Full-energy peak (FEP) efficiency of the CeBrA demonstrator}

Several factors affect the $\gamma$-ray detection efficiency of the CeBrA demonstrator. These include, {\it e.g.}, the crystal size, the distance from the $\gamma$-ray source to the detector (minimum of 4.5 inches), and additional attenuating objects between the source and detector.

\begin{figure}[t]
    \centering
        \includegraphics[width=1\linewidth]{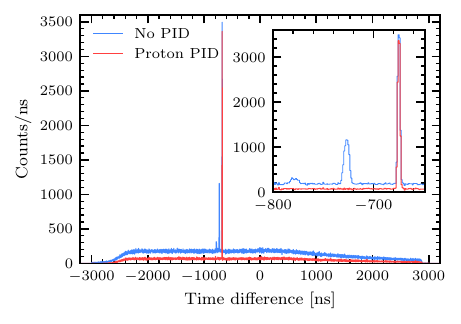}
        \caption{\label{fig:timing}{Example of time-difference spectrum between a $2'' \times 2''$ CeBr$_3$ detector of the CeBrA demonstrator and the scintillator of the SE-SPS focal-plane detector. Clear prompt coincidence peaks are observed, which belong to different particle groups, i.e., $\alpha$ particles, tritons, deuterons and protons (blue histogram). The centroid and FWHM values are as follows: protons (-674\,ns, 2.6\,ns), deuterons (-726\,ns, 5.6\,ns), tritons (-778\,ns, 8.7\,ns), and $\alpha$ particles (-727\,ns, 4.9\,ns). If the particle gate is set around the protons, only the coincidence between the protons and $\gamma$ rays remains (red histogram). Differences are caused by the flight-time difference of the light-ion groups through the spectrograph+detector system.}}
\end{figure}

To determine the absolute full-energy peak (FEP) efficiency of the CeBrA demonstrator, we positioned a \nuc{152}{Eu} standard calibration source of known activity at the center of the CeBrA scattering chamber installed in front of the SE-SPS. A \nuc{56}{Co} source of unknown activity was used to determine the absolute FEP efficiency of the CeBrA demonstrator up to $E_{\gamma} = 3.5$\,MeV by scaling the low-energy data to the absolute efficiencies measured with the \nuc{152}{Eu} source. As an example, the measured FEP efficiencies for one $2'' \times 2''$ and the $3'' \times 4''$ CeBr$_3$ detector of the CeBrA demonstrator are shown in Fig.\,\ref{fig:efficiency}. They were obtained with the second configuration shown in Fig.\,\ref{fig:cebra}. A double exponential function was fitted to the experimental data. As can be seen from the residuals in Fig.\,\ref{fig:efficiency}, the fit describes the data well up to $E_{\gamma} = 3.5$\,MeV. For the two configurations, as shown in Fig.\,\ref{fig:cebra}, a $2'' \times 2$'' detector has a typical FEP efficiency of $\sim 0.2\,\%$ while the $3'' \times 4''$ detector has an FEP efficiency of $\sim 0.7\,\%$ at 1.3 MeV. The five-detector demonstrator array has a combined FEP efficiency of around 1.5\,$\%$ at 1.3 MeV. These quoted values depend of course on the distance from the chamber.

\subsection{\nuc{49}{Ti}$(d,p\gamma)$\nuc{50}{Ti} and \nuc{61}{Ni}$(d,p\gamma)$\nuc{62}{Ni} experiments: Coincidence timing, gating, and $\gamma$-decay intensities}

\begin{figure}[t]
\centering
        \includegraphics[width=\linewidth]{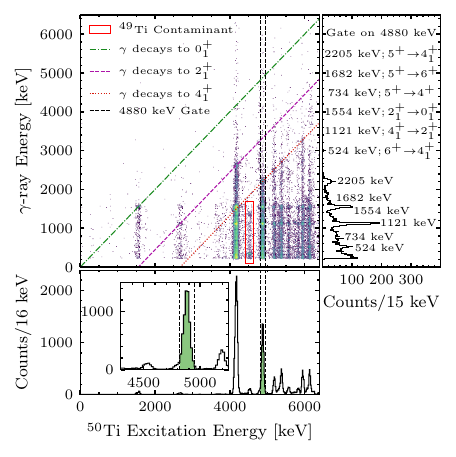}
        \caption{\label{fig:matrix}{Particle-$\gamma$ coincidence matrix for \nuc{49}{Ti}$(d,p\gamma)$\nuc{50}{Ti}. A narrow coincidence condition around the prompt proton-$\gamma$ coincidence peak shown in Fig.\,\ref{fig:timing} was applied. Diagonals corresponding to $\gamma$-ray decays leading to different final excited states can be observed. The intercept with the $x$-axis specifies the excitation energy of that final state. The $y$-projection shows an example for a $\gamma$-ray spectrum when gated on an excitation energy of $E_x = 4880$ keV (vertical gate). Only $\gamma$ rays originating from the corresponding decay cascade are observed.}}
\end{figure}

For the first in-beam particle-$\gamma$ coincidence tests, we used a continuous deuteron beam of beam energy $E_d = 16$\,MeV and set up an external trigger requiring a coincidence between the focal-plane scintillator and the CeBr$_3$ detectors within a time window of $\Delta t = \pm 3$ $\mu$s. The configuration in the top panel of Fig.\,\ref{fig:cebra} was used. With typical currents on target of 6 nA, the CeBr$_3$ detectors were counting at rates of about 200\,kcps, which is possible because of their extremely short signal pulses of $\sim$200\,ns without running into significant pile-up problems (see Fig.\,\ref{fig:signal} for an example). The coincidence timing spectrum is shown in Fig.\,\ref{fig:timing}. Clear prompt coincidence peaks are observed on top of a flat random coincidence background. The three distinct peaks correspond to coincidences between $\gamma$ rays detected with the CeBr$_3$ detectors and light ions detected with the SE-SPS focal-plane detector, respectively. When a particle gate is set around the protons, only the corresponding coincidence peak remains (see Fig.\,\ref{fig:timing}).

\begin{figure}[t!]
\centering
\includegraphics[width=\linewidth]{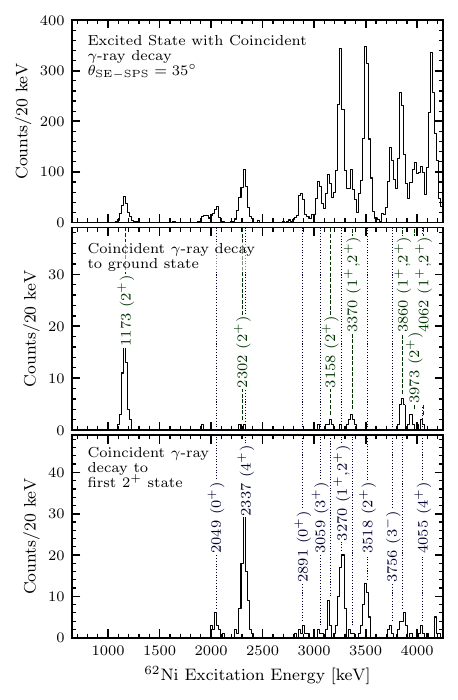}
\caption{\label{fig:Ni}{Particle-$\gamma$ coincidences measured in \nuc{61}{Ni}$(d,p\gamma)$\nuc{62}{Ni}. (Top) Focal-plane (FP) spectrum if any coincident $\gamma$-ray was detected. (Middle) FP spectrum if $\gamma$-ray transition led directly to the ground state. (Bottom) FP spectrum if $\gamma$-ray transition led directly to the $2^+_1$ state. States are marked with their excitation energies and adopted spin-parity assignment. For illustrative purposes, only the energy region between 700 keV and 4200 keV of the focal-plane spectrum is shown. See text for further discussion.}}
\end{figure}

We chose the \nuc{49}{Ti}$(d,p\gamma)$\nuc{50}{Ti} reaction as a test case since the excitation spectrum is comparably simple at low excitation energies, since low-lying excited states are known to decay to different final states, and since \nuc{50}{Ti} had previously been studied in $(d,p\gamma)$\,\cite{Son84a}. The particle-$\gamma$ coincidence matrix for the \nuc{49}{Ti}$(d,p\gamma)$\nuc{50}{Ti} reaction measured with the combined CeBrA+SE-SPS setup, and with the proton gate in the PID plot (see Fig.\,\ref{fig:ede}) as well as a narrow timing gate of 4.5-ns width around the prompt peak (see Fig.\,\ref{fig:timing}) applied is shown in Fig.\,\ref{fig:matrix}. Different features including diagonal bands corresponding to $\gamma$-ray decays leading to different final states can be clearly identified. A \nuc{49}{Ti} contaminant originating from a \nuc{48}{Ti} target contaminant is also identified. Such a contaminant would be missed in particle-singles experiments. A state at the corresponding energy is indeed listed in the NNDC database \cite{ENSDF}. The inset of Fig.\,\ref{fig:matrix} also shows a $\gamma$-ray spectrum obtained when an excitation-energy gate with a width of $\sim 130$\,keV is set around the 4880-keV excited state. This example highlights the power and sensitivity of particle-$\gamma$ coincidences with the SE-SPS and $\gamma$-ray detectors as only $\gamma$ rays originating from the corresponding $\gamma$-decay cascade are observed. The statistics shown here were obtained in 48 hours. During that time, $\sim 69000$ protons were detected with the focal-plane detector for the 4880-keV state. With the prompt timing gate applied, 6400 of these protons were detected in coincidence with $\gamma$ rays. The coincident $\gamma$ rays are shown in Fig.\,\ref{fig:matrix} (see $y$-axis projection). 

For the 4880-keV ($J^{\pi}=5^{+}$) state, $\gamma$-decay intensities of $I_{\gamma} = 2.1(2)$ for the 734-keV, $5^+ \rightarrow 4^+$ transition, 8(2) for the 1682-keV, $5^+ \rightarrow 6^+_1$ transition, and 100(6) for the 2205-keV, $5^+ \rightarrow 4^+_1$ transition are adopted\,\cite{ENSDF}. We determined an upper limit of $I_{\gamma} \leq 15$ for the 734-keV, $5^+ \rightarrow 4^+$ transition, and $\gamma$-decay intensities of 24(7) for the 1682-keV, $5^+ \rightarrow 6^+_1$ transition and 100(11) for the 2205-keV, $5^+ \rightarrow 4^+_1$ transition. As our $\gamma$-decay intensity for the 1682-keV transition deviates from the adopted value, we note that intensities of 22(3) and 14 were reported for this $\gamma$-decay branch following the \nuc{50}{Sc} $\beta^-$ decay\,\cite{Alb84a} and the $(d,p\gamma)$ reaction\,\cite{Son84a}, respectively. These intensities are in agreement with our new data. We also verified that the yield for the 1682-keV, $5^+ \rightarrow 6^+_1$ transition is consistent with the number of counts observed for the 524-keV, $6^+ \rightarrow 4^+_1$ transition when an excitation gate of 4880\,keV is set.

\begin{figure}[t!]
\centering
\includegraphics[width=\linewidth]{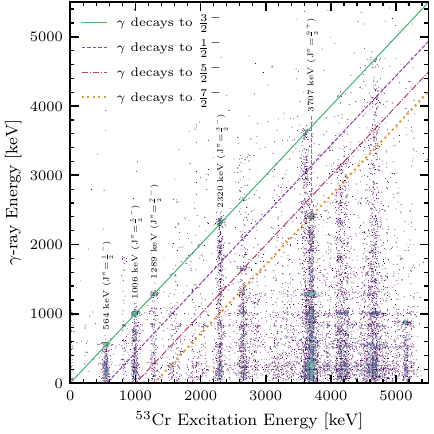}
\caption{\label{fig:Cr}{Particle-$\gamma$ coincidence matrix for \nuc{52}{Cr}$(d,p\gamma)$\nuc{53}{Cr}. Also shown are diagonals on which $\gamma$ decays of excited states leading to specific final states of \nuc{53}{Cr} can be identified. The intercept with the $x$-axis specifies the excitation energy of that final state.}}
\end{figure}

Another example for the power of particle-$\gamma$ coincidences with high-resolution magnetic spectrographs is shown in Fig.\,\ref{fig:Ni}. In this case, diagonal gates (as also shown in Fig.\,\ref{fig:matrix}) either requiring direct $\gamma$-ray decays to the ground or $2^+_1$ state are applied. As can be seen, different excited states populated in \nuc{61}{Ni}$(d,p)$\nuc{62}{Ni} and detected with the focal-plane detector can be picked up this way. While in the ground-state diagonal gate we primarily observe states with spin-parity assignment $J^{\pi} = 1^+, 2^+$, also $J^{\pi} = 3^+$ and $4^+$ states are observed when selecting $\gamma$ decays which lead to the 1173-keV, $J^{\pi} = 2^+_1$ state. This gating technique can provide important complementary information for better constraining spin-parity assignments and identifying which states were populated in a specific nuclear reaction. The \nuc{61}{Ni}$(d,p\gamma)$\nuc{62}{Ni} test experiment ran for roughly 61 hours.

\subsection{\nuc{52}{Cr}$(d,p\gamma)$\nuc{53}{Cr} and \nuc{34}{S}$(d,p\gamma)$\nuc{35}{S} experiments: Proton-$\gamma$ angular correlations and lifetime determination}

For the second set of in-beam particle-$\gamma$ coincidence tests, the CeBr$_3$ detectors were rearranged to be in the same plane ($\phi_{\gamma}$ = 0$\degree$) relative to the beam axis to measure particle-$\gamma$ angular correlations. The $2'' \times 2''$ CeBr$_3$ detectors were placed at $\theta_{\gamma}$ = 96(14)$\degree$, 211(14)$\degree$, 241(14)$\degree$, and 271(10)$\degree$, and the $3'' \times 4''$ CeBr$_3$ detector at $\theta_{\gamma}$ = 141(22)$\degree$ relative to the beam axis. The port from the scattering chamber to the SE-SPS was fixed at $\theta_{\mathrm{SE-SPS}}$ = 37$\degree$. All angles are measured clockwise relative to the beam axis. Also for this set of experiments, a 16\,MeV deuterium beam was used and an external trigger was set up requiring a coincidence between the focal-plane scintillator and the CeBr$_3$ detectors within a time window of $\Delta t = \pm 3$ $\mu$s.

\begin{figure}[!t]
    \centering
    \includegraphics[width=\linewidth]{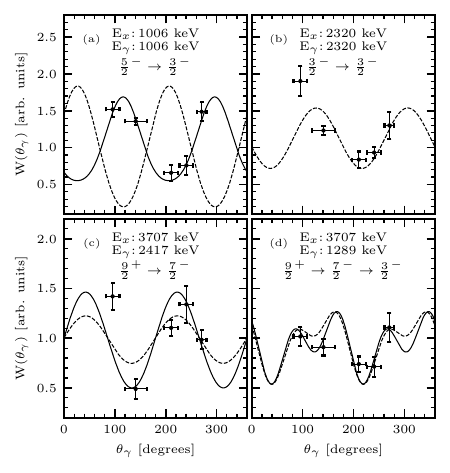}
    \caption{Proton-$\gamma$ angular correlations measured for excited states of \nuc{53}{Cr} via \nuc{52}{Cr}$(d,p\gamma)$\nuc{53}{Cr} (symbols). Excitation energies $E_x$ of the states, the $\gamma$-ray energy $E_{\gamma}$ of the transition, as well as the spin-parity assignments $J^{\pi}$ of the states involved are specified. In (c) and (d), the primary 2417-keV, $9/2^+ \rightarrow 7/2^-$ transition, depopulating the 3707-keV level, as well as the secondary 1290-keV, $7/2^- \rightarrow 3/2^-$ transition are shown, respectively. In addition, predictions from combined ADWA calculations with \textsc{chuck3}\,\cite{CHUCK} yielding scattering amplitudes and \textsc{angcor}\,\cite{angcor} calculations using these scattering amplitudes to generate the angular correlations are shown for each transition (lines). In panels (a)-(d), the dashed lines correspond to predictions when using the adopted multipole mixing ratios\,\cite{ENSDF}, while solid lines used different multipole mixing ratios. See text for further discussion.}
    \label{fig:53Cr_angular_correlation}
\end{figure}

For the \nuc{52}{Cr}$(d,p\gamma)$\nuc{53}{Cr} experiment, a target of 300-$\mu$g/cm$^2$ natural Cr evaporated onto a 20-$\mu$g/cm$^2$ thick backing of natural Carbon was used; \nuc{52}{Cr} has a natural abundance of 83.8\,$\%$.  The particle-$\gamma$ coincidence matrix is shown in Fig.\,\ref{fig:Cr}. By applying gates in the corresponding matrix for the individual CeBrA detectors and, hence, selecting the excited state of interest, the proton-$\gamma$ angular correlations shown in Fig.\,\ref{fig:53Cr_angular_correlation} were obtained. In addition, predictions from combined Adiabatic Distorted Wave Approximation (ADWA) calculations with \textsc{chuck3}\,\cite{CHUCK} yielding scattering amplitudes and \textsc{angcor}\,\cite{angcor} calculations using these scattering amplitudes to generate the angular correlations are shown in Fig.\,\ref{fig:53Cr_angular_correlation}. These predictions (lines) were scaled to data by minimizing the residuals using the experimental uncertainties as weights. All theoretical proton-$\gamma$ angular correlations were averaged over the solid-angle acceptance $\Delta \Omega = 4.6$\,msr of the SE-SPS used in our experiments.

For the 1006-keV, $5/2^-_1 \rightarrow 3/2^-_1$ transition, we found that the experimentally measured angular correlation is best described when using a multipole mixing ratio of $\delta \approx -0.58$. Even though similar in magnitude, the absolute value is different from the previously reported values of $\delta = +0.36(2)$ from \nuc{50}{Ti}$(\alpha,n)$\,\cite{Gul73a} and $\delta = +0.80(6)$ from Coulomb excitation\,\cite{Pat77a}. For completeness, we added the angular correlation predicted with the adopted multipole mixing ratio of $\delta = +0.36(2)$\,\cite{ENSDF} [see dashed line Fig.\,\ref{fig:53Cr_angular_correlation}\,(a)], which clearly does not describe our data. For the 2320-keV, $3/2^-_2 \rightarrow 3/2^-_1$, we kept the adopted multipole mixing ratio of $\delta = -0.11(3)$\,\cite{ENSDF} as it described our data reasonably well. A previous $(d,p\gamma)$ experiments reported a multipole mixing ratio of $\delta = 0.00_{-0.09}^{+0.04}$ for the primary 2417-keV, $9/2^+ \rightarrow 7/2^-_1$ transition from the 3707-keV state\,\cite{Car70a}. We found that our measured angular distribution is best described with a multipole mixing ratio of $\delta \approx 0.27$. Also for this transition, we added the predictions when using the multipole mixing ratio of $\delta = 0.00$ as inferred from the previous $(d,p\gamma)$ experiment\,\cite{Car70a} to Fig.\,\ref{fig:53Cr_angular_correlation}\,(c). We studied the secondary 1290-keV, $7/2^-_1 \rightarrow 3/2^-_1$ transition after the $\gamma$ decay of the 3707-keV level as well. This proton-$\gamma$ angular correlation is shown in Fig.\,\ref{fig:53Cr_angular_correlation}\,(d). It is well described when using a multipole mixing ratio of $\delta \approx 0.17$ [solid line in Fig.\,\ref{fig:53Cr_angular_correlation}\,(d)]. The currently adopted value from a Coulomb-excitation experiment is $\delta = 0.072(6)$\,\cite{ENSDF, Pat77a}, which also leads to a fair description of our data. Note that different from other experiments, we can exclude feeding contributions due to our selective prompt-timing and excitation-energy gates.

\begin{figure}[t!]
    \centering
    \includegraphics[width=\linewidth]{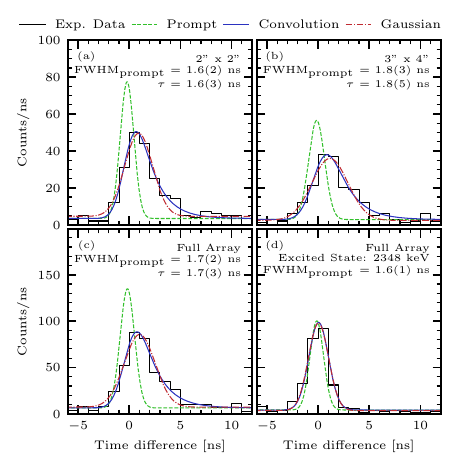}
    \caption{\label{fig:35S_1991_lifetime}{Time-difference spectra of CeBr$_3$ detectors and the scintillator of the SE-SPS focal-plane detector. (a) $2'' \times 2''$ CeBr$_3$ detectors, (b) $3'' \times 4''$ CeBr$_3$ detector, (c) entire CeBrA demonstrator for the 1991-keV state of \nuc{35}{S}, and (d) for the 2348-keV state of \nuc{35}{S}. The lifetime of the 1991-keV state determined using the convolution method and the different detector combinations is shown in panels (a) through (c), respectively. No tail is observed in panel (d) indicative of a short lifetime of the 2348-keV state. See text for more details.}}
\end{figure}

In addition to establishing particle-$\gamma$ coincidences and first proton-$\gamma$ angular correlations with CeBrA and the SE-SPS, we studied the fast-timing capabilities of the combined setup to determine nuclear level lifetimes. The idea is to set a direct excitation-energy gate with the SE-SPS on the state whose lifetime is to be determined, which excludes any feeding. In addition, we do not have to pay a ``$\gamma$-ray detection efficiency penalty'' when choosing a feeding transition to select a specific $\gamma$-decay cascade and when wanting to exclude side-feeding contributions as done in, {\it e.g.}, Ref.\,\cite{Har23a}. Employing the relative timing between the focal-plane scintillator and the CeBrA demonstrator, we determined a lifetime of 1.7(3)\,ns for the 1991-keV state of \nuc{35}{S} after being populated through the \nuc{34}{S}$(d,p)$ reaction [see Fig.\,\ref{fig:35S_1991_lifetime}\,(a)-(c)]. To extract the lifetime, we used the convolution method (see also Fig.\,\ref{fig:lifetime}). The lifetimes determined with the $2'' \times 2''$ detectors and with the $3'' \times 4''$ detector, respectively, are consistent. A Gaussian distribution was fitted to the data to illustrate the asymmetry in the tail, which is caused by the lifetime of the 1991-keV state (see Fig.\,\ref{fig:35S_1991_lifetime}). To further illustrate the influence of a lifetime on the time-difference spectrum, we also studied the 2348-keV state of \nuc{35}{S}. The 2348-keV state has a much shorter lifetime of 1.2(2)\,ps\,\cite{ENSDF}. Consequently, we expect no pronounced tail in the time-difference spectrum. The comparison of Fig.\,\ref{fig:35S_1991_lifetime}\,(d) to Figs.\,\ref{fig:35S_1991_lifetime}\,(a) - (c) clearly showcases that this is the case. We want to point out that our lifetime for the 1991-keV state is in agreement with the adopted value of 1.47(7)\,ns \cite{ENSDF,35S_lifetime} and that our measurement marks only the second measurement of the 1991-keV state's lifetime so far.

\section{Summary and Outlook}

In this article, we presented results on the energy and timing resolution of $2'' \times 2''$ and $3'' \times 4''$ CeBr$_3$ detectors measured with a digital data acquisition system based on the CAEN V1725S digitizer with DPP-PSD firmware. We also reported on the subsequent commissioning experiments with the CeBrA demonstrator for particle-$\gamma$ coincidence experiments at the Super-Enge Split-Pole Spectrograph of the John D. Fox Superconducting Linear Accelerator Laboratory at Florida State University, for which the same data acquisition was used.

The characterization efforts, presented in this article and using standard calibration sources for the measurements, focused on optimizing the energy and timing resolution of the CeBr$_{3}$ detectors. At 1.3\,MeV, we achieved an energy resolution of 2.6\% for the $2'' \times 2''$ detectors and 2.8\% for the $3'' \times 4''$ detector. Using a \nuc{56}{Co} source, we established that the energy resolution of the CeBr$_3$ detectors evolves smoothly and follows the expected $1/\sqrt{E_{\gamma}}$ energy dependence up to $E_{\gamma} = 3.5$\,MeV. The timing resolution varied between 500-590\,ps for the $2'' \times 2''$ CeBr$_{3}$ detectors, while the $3'' \times 4''$ CeBr$_{3}$ detector exhibited a timing resolution of 750\,ps measured relative to a $2'' \times 2''$ detector. For these measurements, the two prompt $\gamma$ transitions emitted in the $\beta^-$ decay of \nuc{60}{Co} were used. We also showed that nuclear level lifetimes in the hundreds of picoseconds range can be measured using the relative timing between the $2'' \times 2''$ CeBr$_3$ detectors. This would be challenging with an array consisting of only $3'' \times 4''$ CeBr$_3$ detectors due to their worse timing resolution. The $2'' \times 2''$ detectors are a reasonable compromise between timing resolution and detection efficiency for $\gamma$ rays with $E_{\gamma} \leq 3$\,MeV compared to even smaller crystal sizes.

To commission the CeBrA demonstrator at the SE-SPS for particle-$\gamma$ coincidence experiments, we performed four experiments; each one specifically chosen to showcase different capabilities of the array. The \nuc{49}{Ti}$(d,p\gamma)$\nuc{50}{Ti} reaction was, for instance, chosen as the excitation spectrum is comparably simple at low excitation energies, as low-lying excited states are known to decay to different final states, and as \nuc{50}{Ti} had previously been studied in $(d,p\gamma)$\,\cite{Son84a}. Besides establishing the excellent coincidence timing between the CeBrA detectors and the SE-SPS, we were able to study the $\gamma$ decay of excited states of \nuc{50}{Ti} to different final states using selective excitation energy gates. As an example, we discussed the $\gamma$ decay of the 4880-keV, $J^{\pi} = 5^+$ state. We showed that the $\gamma$-decay intensities, which we determined with the CeBrA demonstrator, were in excellent agreement with two of the three experiments which had previously studied the $\gamma$ decay of this state. The \nuc{61}{Ni}$(d,p\gamma)$\nuc{62}{Ni} reaction was chosen to highlight the capability of the combined array to select $\gamma$ decays leading to specific final states with so-called diagonal gates and to pick up states with different spin and parity quantum numbers in that way. As an example, we showed that states with $J = 1,2$ can be picked up by selecting $\gamma$ decays leading directly to the $J^{\pi} = 0^+$ ground state of \nuc{62}{Ni} even though several states might be overlapping in a certain excitation-energy range. Using the \nuc{52}{Cr}$(d,p\gamma)$\nuc{53}{Cr} reaction, we proved that very pronounced particle-$\gamma$ angular correlations can already be measured with the limited detection efficiency of the CeBrA demonstrator. By performing combined calculations with the \textsc{chuck3} and \textsc{angcor} computer programs, we furthermore showed that we are sensitive to multipole mixing ratios. To test the fast-timing capabilities of the setup, we chose the \nuc{34}{S}$(d,p\gamma)$\nuc{35}{S} reaction. Using the relative timing between the SE-SPS focal plane scintillator and the CeBr$_3$ detectors and setting a selective excitation energy gate, which excludes feeding, a lifetime of 1.7(3)\,ns was measured for the 1991-keV state. This lifetime is in good agreement with the adopted value of 1.47(7)\,ns and marks only the second measurement of this lifetime so far.

In summary, we commissioned the combined CeBrA and SE-SPS setup for particle-$\gamma$ coincidence experiments. Thanks to the excellent resolution of the SE-SPS, narrow excitation energy gates can be set, which allow the selective study of the $\gamma$ decay of excited states populated through light-ion induced nuclear reactions with the CeBr$_3$ detectors of CeBrA. We demonstrated the capability of the combined setup to measure $\gamma$-decay branching ratios, particle-$\gamma$ angular correlations, and nuclear level lifetimes using a set of carefully chosen $(d,p\gamma)$ test experiments. We also proved that, because of the fast recovery of the CeBr$_3$ signals within a few hundred nanoseconds, $(d,p\gamma)$ experiments with CeBrA at the SE-SPS can be performed with typical currents of several nanoamperes on target, which would be challenging with High-Purity Germanium detectors in the proximity of the SE-SPS due to increased neutron and $\gamma$-ray background. The CeBr$_3$ detectors run at rates of about 250\,kcps with these currents on target and, during and after our experiments, we did not observe any loss in spectral quality. We have already tested that, {\it e.g.,} $(\alpha,d\gamma)$ and $(\nuc{6}{Li},d\gamma)$ experiments can be conducted with currents on target as high as several hundreds of nanoamperes and that the level of background radiation is much lower than in $(d,p\gamma)$. We will continue deploying the CeBrA demonstrator for detailed nuclear structure studies at the SE-SPS. However, we intend to add more detectors to CeBrA in order to increase its $\gamma$-ray detection efficiency and angular coverage in the near future. Specifically, we would like to add more large-volume CeBr$_3$ crystals to increase the $\gamma$-ray detection efficiency at energies as high as those needed to study the $\gamma$ decay of excited states up to the particle-separation thresholds. Such an upgrade would also enable particle-$\gamma$ coincidence experiments for reactions with comparably small cross sections in a reasonable amount of time.

\section*{Acknowledgements}
This work was supported by the National Science Foundation (NSF) under Grant No. PHY-2012522 (WoU-MMA: Studies of Nuclear Structure and Nuclear Astrophysics) and by Florida State University. We want to express our gratitude to Alexandra Gade and Dirk Weisshaar for lending us their LaBr$_3$(Ce) detectors. We also want to explicitly thank J. Aragon, P. Barber, R. Boisseau, B. Schmidt, R. Smith, and CAEN’s support staff for their support prior, during, and after our commissioning experiments. M.S. also wants to thank K.W. Kemper for many inspiring discussions. Part of the data was obtained during the 2023 Research Experience for Undergraduate Students (REU) at the John D. Fox Laboratory. The \nuc{34}{S}, \nuc{47,49}{Ti}, and \nuc{61}{Ni} targets were provided by the Center for Accelerator Target Science at Argonne National Laboratory.

\bibliographystyle{apsrev4-1}
\bibliography{CeBrA_NIM_A.bib}

\end{document}